\documentclass[prl,aps,groupedaddress,twocolumn,floatfix,nofootinbib]{revtex4}
\usepackage{amsmath,amsthm,amsfonts,amssymb,graphics,graphicx,epsfig,color,times,natbib,subfigure}
\usepackage{bbm} 

\newcommand{\ket}[1]{\mbox{$ | #1 \rangle $}}
\newcommand{\bra}[1]{\mbox{$ \langle #1 | $}}

\newcommand{\be}{\begin{equation}}
\newcommand{\ee}{\end{equation}}
\newcommand{\ba}{\begin{eqnarray}}
\newcommand{\ea}{\end{eqnarray}}

\newcommand{\one}{\leavevmode\hbox{\small1\normalsize\kern-.33em1}}

\def\sign{\mathrm{sign}}

\begin{document}

\title{The local content of bipartite qubit correlations}
\author{Cyril Branciard$^1$, Nicolas Gisin$^1$, Valerio Scarani$^2$}
\affiliation{$^1$ Group of Applied Physics, University of Geneva, Geneva, Switzerland\\ $^2$ Centre for Quantum Technologies and Department of Physics, National University of Singapore, Singapore}
\date{\today}

\begin{abstract}
One of the last open problems concerning two qubits in a pure state is to find the exact local content of their correlation, in the sense of Elitzur, Popescu and Rohrlich (EPR2) [Phys. Lett. A 162, 25 (1992)]. We propose a new EPR2 decomposition that allows us to prove, for a wide range of states $\ket{\psi(\theta)}=\cos\theta\ket{00}+\sin\theta\ket{11}$, that their local content is $\bar{p_L}(\theta) = \cos 2\theta$. We also share reflections on how to possibly extend our result to all two-qubit pure states.
\end{abstract}

\maketitle

\section{Introduction}

The incompatibility of quantum mechanics with local variable theories, as shown by Bell~\cite{bell64}, lies at the statistical level: local variable theories cannot reproduce all statistical predictions of quantum theory. In a typical Bell experiment, one observes correlations between the measurement results of two partners (Alice and Bob), and averages them over measurements on many pairs of particles. One may then conclude that non-locality was observed if the average correlations thus obtained violate a Bell inequality.

However, if the statistics of the observations exhibit non-locality, it does not imply that all individual pairs behave non-locally. This observation lead Elitzur, Popescu and Rohrlich~\cite{elitzur_quantum_1992}, hereafter referred to as EPR2, to wonder whether one could consider that a fraction of the pairs still behaves locally, while another fraction would behave non-locally (and possibly more non-locally than quantum mechanics allows).

More explicitly, writing $P_Q$ the quantum mechanical probability distribution for Alice and Bob's results,
the EPR2 approach consists in decomposing $P_Q$ as a convex sum of a local part, $P_L$, and of a non-local part, $P_{NL}$, in the form \ba P_Q = p_L P_L + (1-p_L) P_{NL}, \quad \textrm{with } p_L \in [0,1] . \label{epr2_decomp} \ea
The maximal weight $\bar{p_L} = \max p_L$ that can be attributed to the local part can be regarded as a measure of (non-)locality of the quantum distribution $P_Q$. Finding this maximal possible local weight is not a trivial problem; so far one only knows how to calculate lower and upper bounds on $\bar{p_L}$.

In this paper, we concentrate on the simplest case of a quantum probability distribution originating from Von Neumann measurements on two-qubit pure states. After recalling previously known results for this case, we propose a new EPR2 decomposition and derive a new lower bound on $\bar{p_L}$, which reaches the previously known upper bound~\cite{scarani_local_2008} for a wide class of states. This gives a definite value for the exact local content $\bar{p_L}$ of those states. We then share reflections on how one might possibly extend our result to all two-qubit pure states.

\section{The EPR2 approach for two-qubit pure states}

\subsection{Correlations of two-qubit pure states}

Without loss of generality, any two-qubit pure state can be written in the form \ba \ket{\psi(\theta)} = \cos \theta \ket{00} + \sin \theta \ket{11} \label{def_state} \ea with $\theta \in [0,\frac{\pi}{4}]$. In the following, we shall use the notation $c = \cos 2\theta, s = \sin 2\theta$ (with $c,s \in [0,1]$).

Each qubit is subjected to a Von Neumann measurement, labeled by unit vectors $\vec a$ and $\vec b$ on the Bloch sphere $S^2$. Let us denote by $a_z$ and $b_z$ the $z$ components of $\vec a$ and $\vec b$, by $a_\perp=\sqrt{1-a_z^2}$ and $b_\perp=\sqrt{1-b_z^2}$ the amplitudes of the components of $\vec a$ and $\vec b$ in the $xy$ plane, and by $\chi\in \,]\!-\pi,\pi]$ the difference between the azimuthal angles of $\vec a$ and $\vec b'$, with $\vec b'$ defined as the reflection of $\vec b$ with respect to the $xz$ plane\footnote{The axes $x,y,z$ of the Bloch sphere are defined as usual: $\ket{0}$ and $\ket{1}$ are identified with the north and south poles (i.e., along the $z$ axis), while the state $(\ket{0}+\ket{1})/\sqrt{2}$ defines the $x$ direction. We introduce $\vec b'=(b_x,-b_y,b_z)$ to account for the minus sign in front of $a_yb_y$ in (\ref{EQ}).}.

With these notations, and for binary results $\alpha, \beta = \pm 1$, quantum mechanics predicts the following conditional probability distribution:
\ba \begin{array}{rcl}
P_Q(\alpha,\beta|\,\vec a, \vec b) &=& \frac{1}{4} \big(1 + \alpha \, M_Q(\vec a) + \beta \, M_Q(\vec b) \\ && \qquad \quad + \ \alpha \beta \, E_Q(\vec a,\vec b)\big)
\end{array} \label{PQ} \ea
\ba
& \mathrm{with} \quad M_Q(\vec a) = c \, a_z , \quad M_Q(\vec b) = c \, b_z , \label{MQ} \\
& \begin{array}{rl}
& E_Q(\vec a, \vec b) = a_z b_z + s \, (a_x b_x - a_y b_y) \\ & \phantom{E_Q(\vec a, \vec b)} = a_z b_z + s \, a_\perp b_\perp \cos\chi \, . \end{array} \label{EQ} \ea

As explained before, the EPR2 problem is to find a decomposition of $P_Q(\alpha,\beta|\,\vec a, \vec b)$ as a convex sum of a local and a non-local probability distribution, in the form (\ref{epr2_decomp}). For a given state (i.e., a given value of $\theta$), the equality is required to hold for all possible measurements $\vec a,\vec b$ and for all results $\alpha, \beta$. The weight $p_L \in [0,1]$ of the local distribution should be independent of the measurements and of the outcomes.

The probability distribution $P_L(\alpha,\beta|\,\vec a, \vec b)$ is required to be local, in the sense that it can be explained by local variables $\lambda$, i.e. it can be decomposed in the form \ba P_L(\alpha,\beta|\,\vec a, \vec b) = \int \mathrm{d} \! \lambda \, \rho(\lambda) \ P^A_\lambda(\alpha|\vec{a})P^B_\lambda(\beta|\vec{b}) . \ea On the other hand, no restriction is imposed on $P_{NL}$, except that it must be non-negative for all inputs and outputs: \ba P_{NL} = \frac{1}{1-p_L}(P_Q-p_L P_L) \, \geq \, 0 \ . \label{PNL_pos} \ea In particular, $P_{NL}$ is allowed to be even more non-local than quantum mechanical correlations\footnote{Note however, that $P_{NL} = \frac{1}{1-p_L}(P_Q-p_L P_L)$ is by construction non-signaling.}.

The goal is to find, for a given state, a decomposition with the largest possible value for $p_L$, denoted $\bar{p_L}(\theta)$, which characterizes the locality of the probability distribution $P_Q$ as defined by (\ref{PQ}--\ref{EQ}).


\subsection{Previously known results and conjecture}

In their original paper~\cite{elitzur_quantum_1992}, Elitzur, Popescu and Rohrlich proposed an explicit local probability distribution $P_L$, which lead to an EPR2 decomposition with $p_L = \frac{1-s}{4}$. This was the first known lower bound on $\bar{p_L}(\theta)$. Clearly, this was not optimal, at least when approaching the product state ($\theta = 0$, i.e., $s = 0$) which is fully local, and therefore satisfies $\bar{p_L}(0) = 1$.

They also argued that for the maximally entangled state ($\theta = \pi/4$), $P_Q$ contains no local part: $\bar{p_L}(\frac{\pi}{4}) = 0$, i.e. no EPR2 decomposition with $p_L > 0$ exists for this state. This is in fact a much more general result, as shown later by Barrett \emph{et al}~\cite{barrett_maximally_2006}: the maximally entangled state of two $d$-dimensional quantum systems, for any dimension $d$, has no local component.

\bigskip

In~\cite{scarani_local_2008}, one of the authors could improve on the first lower bound for $\bar{p_L}(\theta)$, as he gave an explicit decomposition that achieves $p_L = 1-s$. Interestingly, it was noted that this is the largest possible value that can be attributed to $p_L$, if $P_L$ depends only on the $z$ components $a_z$ and $b_z$ of $\vec a$ and $\vec b$. Recently, this bound ($\bar{p_L} \geq 1-s$) has been extended to mixed two-qubit states by noticing that $s$ is actually the concurrence of the state~\cite{zhang_local_2009}.

It is worth mentioning here that finding lower bounds on $\bar{p_L}$ (ie., explicit EPR2 decompositions) can be useful for the problem of simulating quantum correlations with non-local resources, as only the non-local part then needs to be simulated. The decomposition of~\cite{scarani_local_2008} was thus successfully used to simulate partially entangled 2-qubit states~\cite{brunner_simulation_2008}.

\bigskip

On the other hand, an upper bound on $\bar{p_L}(\theta)$ can be obtained with the help of Bell inequalities~\cite{barrett_maximally_2006}. Let $I \leq I_L$ be a Bell inequality (defined by a linear combination of conditional probabilities), $I_Q$ the quantum value obtainable with the probability distribution $P_Q$, and $I_{NS} (> I_L)$ the maximum value obtainable with non-signaling distributions. Then from (\ref{epr2_decomp}) it follows that $I_Q \leq p_L I_L + (1-p_L) I_{NS}$, i.e., \ba p_L \leq \frac{I_{NS}-I_Q}{I_{NS}-I_L} \ . \ea Using the family of ``chained Bell inequalities"~\cite{pearle_hidden-variable_1970,braunstein_wringing_1990}, an upper bound for $\bar{p_L}(\theta)$ was derived (numerically) in~\cite{scarani_local_2008}, namely $\bar{p_L}(\theta) \leq \cos2\theta$.

\bigskip

So far, the gap was still open between the two bounds \ba 1-\sin 2\theta \leq \bar{p_L}(\theta) \leq \cos 2\theta \ . \ea It has been conjectured~\cite{scarani_elitzur-popescu-rohrlich_2007} that there should exist an EPR2 decomposition that reaches the upper bound, i.e. with $p_L = c$. If this could be proven to be true, then the lower and upper bounds would coincide, and one could conclude that the value of $\bar{p_L}(\theta)$ is exactly $\cos 2\theta$.

In the following we describe our (partially successful) attempts to prove this conjecture.

\section{Reformulation of the problem \qquad \qquad \qquad \qquad to prove the conjecture}

Our goal is now to see whether it is indeed possible to attribute a weight $p_L = c$ in the EPR2 decomposition of the 2-qubit probability distribution $P_Q$ (\ref{PQ}), and write \ba P_Q = c P_L + (1-c) P_{NL} \ . \label{decomp_pl_c} \ea For that, we want to find an explicit local probability distribution $P_L$, such that $P_{NL} = \frac{1}{1-c}(P_Q-cP_L)$ is a valid probability distribution, i.e. $P_{NL}$ must be non-negative. The problem thus translates into \ba \mathbf{Problem :} \quad \mathrm{ find} \ P_L, \ \mathrm {such \ that} \ \ P_Q - c P_L \geq 0 \ . \ea

\bigskip

At this point, we shall impose an additional (and possibly questionable) constraint on the EPR2 decomposition we are looking for: we want the non-local part to have random marginals\footnote{Note that this constraint precisely justifies the choice $p_L = c$. Indeed, if one can find an EPR2 decomposition with random non-local marginals, then for the setting $\vec z$, $M_Q(\vec z) = c = p_L M_L(\vec z)$, which implies $p_L \geq c$. Now, $c$ is known to be an upper bound for $p_L$, and therefore $p_L = c$.}, i.e., with obvious notations, $M_{NL}(\vec a) = M_{NL}(\vec b) = 0$. The intuition is that the marginals are local properties, which should be concentrated on the local component only\footnote{However, one hint that the argument is questionable is the fact that the no-signaling polytopes for arbitrarily many measurements but binary outcomes contain extremal points with non-random marginals~\cite{barrett_popescu-rohrlich_2005,jones_interconversion_2005}.}.

As equality (\ref{decomp_pl_c}) should also hold individually for the marginals on Alice's and Bob's sides, one should then have $M_Q(\vec a) = c M_L(\vec a)$ and $M_Q(\vec b) = c M_L(\vec b)$, i.e. \ba M_L(\vec a) = a_z \ , \quad M_L(\vec b) = b_z \ . \ea
With these constraints, the condition $P_Q - c P_L \geq 0$ reads: \ba &&
\mathrm{for \ all} \ \alpha, \beta, \vec a, \vec b, \nonumber \\ && \qquad 1 - c +
\alpha \beta (E_Q(\vec a, \vec b) - c E_L(\vec a, \vec b)) \geq 0 \ .
\ea

Thus, the problem now translates into: \ba & \mathbf{Problem :} & \mathrm{ find} \ P_L, \ \mathrm {such \ that} \nonumber \\ && \ \left\{ \begin{array}{l} M_L(\vec a) = a_z \\ M_L(\vec b) = b_z \\ |E_Q(\vec a, \vec b) - c E_L(\vec a, \vec b)| \leq 1-c \end{array} \right. \label{new_problem} \ea

\section{Proposal for a new EPR2 decomposition}

As we are dealing with qubits, the natural geometry of the problem involves unit vectors on the Bloch sphere; we shall propose a local component $P_L$ that makes the most of this geometry. Inspired also by models that Bell devised to reproduce the measurement statistics on a single qubit in the state $\ket{0}$ ~\cite{bell_problem_1966} (which gives precisely the marginals we want), or to approximate the statistics of the singlet state~\cite{bell64}, we introduce the following model to define $P_L$:

\bigskip

{\bf Local model: } \emph{Alice and Bob share a random local variable $\vec \lambda$, uniformly distributed on the Bloch sphere. When Alice receives the measurement direction $\vec a$, she outputs $\alpha(\vec a,\vec\lambda) = \sign(a_z - \vec a \cdot \vec \lambda)$. Similarly, when Bob receives the measurement direction $\vec b$, he outputs $\beta(\vec b,\vec\lambda) = \sign(b_z - \vec b' \cdot \vec \lambda)$, where $\vec b'$ is the reflection of $\vec b$ with respect to the $xz$ plane.}

\bigskip

Let us check whether the constraints (\ref{new_problem}) are satisfied.

\bigskip

\paragraph{Marginals.}

Alice's and Bob's marginals corresponding to our local probability distribution $P_L$ are, as required in (\ref{new_problem}): \ba && M_L(\vec a) = \int \!\!\!\!\! \int_{S^2} \frac{\mathrm{d} \! \lambda}{4\pi} \ \sign(a_z - \vec a \cdot \vec \lambda) \ = \ a_z \\ && M_L(\vec b) = \int \!\!\!\!\! \int_{S^2} \frac{\mathrm{d} \! \lambda}{4\pi} \ \sign(b_z - \vec b' \cdot \vec \lambda) \ = \ b_z \ . \ea

\bigskip

\paragraph{Correlation term.}

The details for the calculation of the local correlation coefficient $E_L(\vec a, \vec b)$ are given in Appendix A. We find
\ba E_L(\vec a, \vec b) &=& \int \!\!\!\!\! \int_{S^2} \frac{\mathrm{d} \! \lambda}{4\pi} \ \sign(a_z - \vec a \cdot \vec \lambda) \ \sign(b_z - \vec b' \cdot \vec \lambda) \nonumber \\ \label{EL} \\
&=& \left\{ \begin{array}{l}
1 - |a_z-b_z| \qquad \mathrm{if }\ \chi = 0 \ , \vspace{3mm} \\ |a_z + b_z| - 1 \qquad \mathrm{if }\ \chi = \pi \ , \vspace{3mm} \\ \begin{array}{c}1 - \frac{2|\chi|}{\pi} + \frac{2}{\pi}a_z \arctan(\frac{a_\perp b_z - a_z b_\perp \cos \chi}{b_\perp \sin |\chi|}) \\ + \frac{2}{\pi}b_z \arctan(\frac{a_z b_\perp - a_\perp b_z \cos \chi}{a_\perp \sin |\chi|}) \end{array} \\ \phantom{1 - |a_z-b_z|} \qquad \mathrm{if }\ 0 < |\chi| < \pi \ .
\end{array} \right.
\label{EL_chi} \ea

\bigskip

One can then check\footnote{This can be proven analytically for the cases when $\chi = 0$ or $\pi$: for a given $c$, one gets the result by looking at the maximum of the function $|E_Q-cE_L|$ for all $a_z, b_z$. For $0 < |\chi| < \pi$ on the other hand, we checked the bound (\ref{bnd_EQcEL}) numerically; for each value of $c$, it was only 3 parameters to vary, so we are confident that the numerics are trustworthy.} that for all settings $\vec a$ and $\vec b$, \ba |E_Q(\vec a, \vec b) - c E_L(\vec a, \vec b)| \, \leq \, \max(1-c,c-s) \ . \label{bnd_EQcEL}\ea

The last constraint in (\ref{new_problem}) is thus satisfied when $c-s \leq 1-c$, i.e. when $c \leq \frac{4}{5}$:
\ba && \mathrm{For \ all} \ c \leq 0.8, \ \mathrm{for \ all \ } \vec a, \vec b, \nonumber \\ && \qquad \qquad |E_Q(\vec a, \vec b) - c E_L(\vec a, \vec b)| \leq 1-c \ . \ea

\section{Conclusion regarding our EPR2 decomposition}

When $c \leq 0.8$, since our local probability distribution $P_L$ satisfies the three constraints (\ref{new_problem}), it defines a valid EPR2 decomposition for $P_Q$, with a local weight that can take the value $p_L = c$. This gives the lower bound $\bar{p_L}(\theta) \geq \cos 2\theta$ for all pure two-qubit states (\ref{def_state}) such that $\cos2\theta \leq 0.8$ (or $\theta \gtrsim 0.1\pi$). As $\cos2\theta$ is also known to be an upper bound for $\bar{p_L}(\theta)$~\cite{scarani_local_2008}, we conclude that this is actually its definite value: \ba \mathrm{when} \ \cos2\theta \leq 0.8, \quad \bar{p_L}(\theta) = \cos 2 \theta \ . \ea

When $\cos 2\theta > 0.8$ however, there exists measurement settings $\vec a, \vec b$ for which the third constraint in (\ref{new_problem}) is not satisfied by $P_L$\footnote{Take $\vec a = - \vec b = \vec x$ for instance: $E_Q(\vec x,-\vec x)-cE_L(\vec x,-\vec x) = c-s > 1-c$ if $c > 0.8$.}. Our local probability distribution cannot be attributed a weight $p_L=c$ in that case.

Still, our decomposition gives a non-trivial lower bound on $p_L$ even when $c > 0.8$, namely\footnote{The lower bound is given by $\min_{\alpha,\beta,\vec a,\vec b} \frac{P_Q}{P_L}$. As for the case $c \leq 0.8$, the bound can be obtained analytically for $\chi = 0$ or $\pi$, and was checked numerically for $0<|\chi|<\pi$.} $p_L \geq c+s-1+\sqrt{2(1-c)(1-s)}$. As long as $c \leq \frac{12}{13}$ (or $\theta \gtrsim 0.06\pi$), this lower bound is larger than the previously known bound $1-s$~\cite{scarani_local_2008}, but when $c \geq \frac{12}{13}$ our new decomposition gives a smaller bound.

\begin{figure}
\begin{center}
\includegraphics[width=0.8\columnwidth]{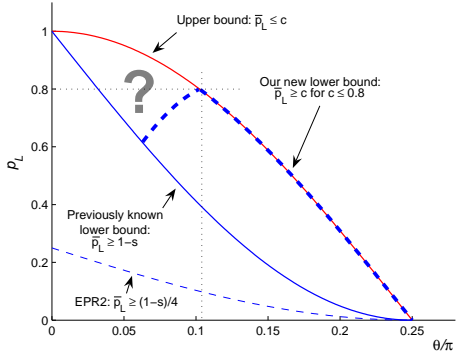}
\caption{Bounds on $\bar{p_L}(\theta)$ in the EPR2 decomposition for two-qubit pure states, as a function of the state parameter $\theta$. Our new lower bound $\bar{p_L}(\theta) \geq \cos 2\theta$, valid for all states such that $\cos 2\theta \leq 0.8$, coincides with the upper bound obtained in~\cite{scarani_local_2008}. There is still a gap between the lower and upper bound when $\cos 2\theta > 0.8$.} \label{fig_epr2_new_bounds}
\end{center}
\end{figure}

Figure~\ref{fig_epr2_new_bounds} summarizes all the bounds we now know on $\bar{p_L}(\theta)$.

\section{Prospects}

We thus could prove the conjecture that $\bar{p_L}(\theta) = \cos 2\theta$ for all states such that $\cos 2\theta \leq 0.8$. This reinforces our opinion, that the result should indeed hold for all pure two-qubit states.

Unfortunately, we could not find so far an EPR2 decomposition with $p_L = \cos 2\theta$ for the very partially entangled states (such that $\cos 2\theta > 0.8$). Let us however share a few reflections on how one could possibly look for a suitable local component $P_L$, which would allow one to prove the conjecture in full generality.

\bigskip

We realize that our local distribution $P_L$ above fails to satisfy the constraints (\ref{new_problem}) when the state under consideration becomes less and less entangled. In our local model, it might be that we correlated the two parties too strongly, by imposing that they share the same local variable $\vec\lambda$.

One idea would be to provide the two parties with two local variables $\vec\lambda_a$ and $\vec\lambda_b \in S^2$, while still considering response functions of the form $\alpha(\vec a,\vec\lambda_a)=\sign(a_z - \vec a \cdot \vec \lambda_a)$ and $\beta(\vec b,\vec\lambda_b)=\sign(b_z - \vec b' \cdot \vec \lambda_b)$. Instead of imposing $\vec\lambda_a = \vec\lambda_b$ as in our previous model, we would correlate $\vec\lambda_a$ and $\vec\lambda_b$ in a smoother way, depending on the state we consider\footnote{To prove the conjecture for $\theta\to 0$, it is actually necessary to have a local part that depends on the state, contrary to our first proposal. Indeed, in the first order in $\theta$ (or $s$), the constraint $|E_Q-cE_L|\leq1-c$ implies that $E_L = E_Q+o(\theta) = a_z b_z + s a_\perp b_\perp \cos\chi + o(s)$.}. In the extreme cases, we would still impose $\vec\lambda_a = \vec\lambda_b$ for the maximally entangled state ($\theta=\frac{\pi}{4}$), while the two $\vec\lambda$'s would be completely decorrelated for the product state ($\theta=0$).

The problem is now to find the proper way to correlate the two $\vec\lambda$'s for each state, i.e. determine the distribution functions $\rho_\theta(\vec \lambda_a,\vec \lambda_b)$. Here are a few properties that we might want to impose on $\rho_\theta(\vec \lambda_a,\vec \lambda_b)$:
\begin{itemize}
\item Forgetting about $\vec\lambda_b$, $\vec\lambda_a$ should be uniformly distributed, and {\em vice versa}. This will ensure in particular that the marginals are those expected: $M_L(\vec a) = a_z, M_L(\vec b) = b_z$. One should thus have:\ba {\textrm{for all}} \ \vec\lambda_a,  \int\!\!\!\!\!\int_{S^2} \!\mathrm{d} \! \lambda_b \rho_\theta(\vec\lambda_a,\vec\lambda_b) = \frac{1}{4\pi} \ ; \\ {\textrm{for all}} \ \vec\lambda_b,  \int\!\!\!\!\!\int_{S^2} \!\mathrm{d} \! \lambda_a \rho_\theta(\vec\lambda_a,\vec\lambda_b) = \frac{1}{4\pi} \ . \label{constr_lambdab_unif} \ea
\item Let us denote by ($\vartheta_{a(b)}$,$\varphi_{a(b)}$) the spherical coordinates of $\vec\lambda_{a(b)}$. It looks very natural to impose that $\rho_\theta(\vec\lambda_a,\vec\lambda_b)$ should only depend on $\vartheta_a,\vartheta_b$ and $\varphi = \varphi_b-\varphi_a$, that it should be symmetrical when exchanging $\vec\lambda_a$ and $\vec\lambda_b$, and that it should have an even dependence on $\varphi$: \ba \rho_\theta(\vec\lambda_a,\vec\lambda_b) &=& \rho_\theta(\vartheta_a,\vartheta_b,\varphi) \\ &=& \rho_\theta(\vartheta_b,\vartheta_a,\varphi) \\ &=& \rho_\theta(\vartheta_a,\vartheta_b,-\varphi)  \ . \label{constr_sym} \ea
\item According to an argument presented in Appendix B, not all pairs $(\vec \lambda_a,\vec \lambda_b)$ should be allowed. More precisely, writing $c_a = \cos\frac{\vartheta_a}{2}, s_a = \sin\frac{\vartheta_a}{2}, c_b = \cos\frac{\vartheta_b}{2}, s_b = \sin\frac{\vartheta_b}{2}$ and $c_\varphi=\cos \varphi$, one should have \ba \rho_\theta(\vec \lambda_a,\vec \lambda_b) = 0 \quad \textrm{if} \quad \frac{s_a s_b}{1-c_a c_b c_\varphi} < s \ . \label{constr_x} \ea
\end{itemize}

\bigskip

We therefore suggest the following research program, to prove the above conjecture for all states: find candidate functions $\rho_\theta(\vec \lambda_a,\vec \lambda_b)$ that have the previous desired properties (\ref{constr_lambdab_unif}--\ref{constr_x}), and then check whether the induced local probability distributions $P_L$ satisfy the constraints (\ref{new_problem}). If one can find such solutions, then this will prove that $\bar{p_L}(\theta) = \cos 2 \theta$ indeed holds for all two-qubit pure states. On the other hand, if it turned out to be impossible to find such a function, then we might need to change our local model, and maybe relax the assumption that the non-local part should have random marginals.

\subsection{Acknowledgements}

We are grateful to Antonio Ac\'in, Mafalda Almeida, Jean-Daniel Bancal, Nicolas Brunner, Loren Coquille, Marc-Andr\'e Dupertuis, Alexandre F\^ete and Stefano Pironio for stimulating discussions.

This work is supported by the Swiss NCCR \emph{Quantum Photonics}, the European ERC-AG \emph{QORE}, and the National Research Foundation and Ministry of Education, Singapore.

\bibliography{bib_EPR2}

\bigskip

\subsection*{Appendix A: Calculation of $E_L(\vec a,\vec b)$}
\label{App_EL}

Here we calculate the correlation coefficient $E_L(\vec a, \vec b)$ for our local probability distribution $P_L$: \ba && E_L(\vec a, \vec b) \nonumber \\ && \quad = \int \!\!\!\!\! \int_{S^2} \frac{\mathrm{d} \! \lambda}{4\pi} \ \sign(a_z - \vec a \cdot \vec \lambda) \ \sign(b_z - \vec b' \cdot \vec \lambda) \nonumber \\ && \quad = \int \!\!\!\!\! \int_{S^2} \frac{\mathrm{d} \! \lambda}{4\pi} \ \big(1 - 2[\vec a \cdot \vec \lambda \geq a_z] \big) \ \big(1 - 2[\vec b' \cdot \vec \lambda \geq b_z] \big) \nonumber \\ && \quad = a_z + b_z - 1 + \frac{1}{\pi} \int \!\!\!\!\! \int_{S^2} \mathrm{d} \! \lambda \ [\vec a \cdot \vec \lambda \geq a_z] \ [\vec b' \cdot \vec \lambda \geq b_z]  , \nonumber \\ \label{EL_int} \ea where $[\cdot]$ is the logical value of what is inside the brakets.
The integral represents the area of the intersection of two spherical caps centered around $\vec a$ and $\vec b'$, and tangent to the north pole of the Bloch sphere.

Let us parameterize $\vec \lambda \in S^2$ by its zenithal and azimuthal angles $(\vartheta, \varphi)$, where $\varphi$ is defined (for simplicity) with respect to the vertical half-plane that contains $\vec a$. As $E_L(\vec a, \vec b)$ should not depend on the sign of $\chi$ (the difference between the azimuthal angles of $\vec a$ and $\vec b'$), it is sufficient to calculate it for $\chi \geq 0$, and simply replace $\chi$ by $|\chi|$ in the final expression. Also, we assume for now that $\vec a$ and $\vec b$ are both in the north hemisphere of the sphere. 

The two spherical caps can then be defined as
\ba && \{ \vec \lambda \ | \ \vec a \cdot \vec \lambda \geq a_z \} \nonumber \\ && \quad = \{ (\vartheta, \varphi) \ | \ \varphi \in [-\frac{\pi}{2},\frac{\pi}{2}] \ \mathrm{and} \ \vartheta \in [0,\vartheta_m^A(\varphi)] \} \nonumber \\ && \{ \vec \lambda \ | \ \vec b' \cdot \vec \lambda \geq b_z \} \nonumber \\ && \quad = \{ (\vartheta, \varphi) \ | \ \varphi \in [\chi-\frac{\pi}{2},\chi+\frac{\pi}{2}] \ \mathrm{and} \ \vartheta \in [0,\vartheta_m^B(\varphi)] \} \nonumber \ea with $\vartheta_m^A(\varphi), \vartheta_m^B(\varphi) \in [0,\pi]$ such that
\ba \cos \vartheta_m^A(\varphi) &=& \frac{a_z^2-a_\perp^2\cos^2\varphi}{a_z^2+a_\perp^2\cos^2\varphi} \ , \nonumber \\ \cos \vartheta_m^B(\varphi) &=& \frac{b_z^2-b_\perp^2\cos^2(\varphi-\chi)}{b_z^2+b_\perp^2\cos^2(\varphi-\chi)} \ . \nonumber \ea

Let us define $\varphi_0$ as the azimuthal angle for which $\vartheta_m^A(\varphi_0)=\vartheta_m^B(\varphi_0)$. The integral in (\ref{EL_int}) can then be calculated as follows: \ba && \int \!\!\!\!\! \int_{S^2} \mathrm{d} \! \lambda \ [\vec a \cdot \vec \lambda \geq a_z] \ [\vec b' \cdot \vec \lambda \geq b_z] \nonumber \\ && \qquad = \int_{\chi-\pi/2}^{\pi/2} \mathrm{d} \varphi \int_0^{\min(\vartheta_m^A(\varphi),\vartheta_m^B(\varphi))} \sin\vartheta \mathrm{d}\vartheta \nonumber \\ && \qquad = \int_{\chi-\pi/2}^{\varphi_0} \mathrm{d} \varphi \big[ 1 - \cos \vartheta_m^B(\varphi) \big] \nonumber \\ && \qquad \quad + \int_{\varphi_0}^{\pi/2} \mathrm{d} \varphi \big[ 1 - \cos \vartheta_m^A(\varphi) \big] \ . \nonumber \ea
Using the antiderivative $\int \! \mathrm{d} \varphi \frac{a_z^2-a_\perp^2 \cos^2\varphi}{a_z^2+a_\perp^2 \cos^2\varphi} = 2 a_z \arctan(a_z \tan\varphi) - \varphi$, we find: \ba && \int \!\!\!\!\! \int_{S^2} \mathrm{d} \! \lambda \ [\vec a \cdot \vec \lambda \geq a_z] \ [\vec b' \cdot \vec \lambda \geq b_z] \nonumber \\ && \qquad\qquad = 2\pi - 2\chi + 2a_z \arctan(a_z\tan\varphi_0) - \pi a_z \nonumber \\ && \qquad\qquad\qquad + 2 b_z \arctan(b_z\tan(\chi-\varphi_0)) - \pi b_z \ . \nonumber \\ \label{intercalottes0} \ea

We note that $\vartheta_m^A(\varphi_0)=\vartheta_m^B(\varphi_0)$ implies $a_z b_\perp \cos(\chi-\varphi_0) = a_\perp b_z \cos \varphi_0$, which in turn implies $\ a_z \tan \varphi_0 = \frac{a_\perp b_z - a_z b_\perp \cos \chi}{b_\perp \sin \chi} \ $ and $\ b_z \tan (\chi-\varphi_0) = \frac{a_z b_\perp - a_\perp b_z \cos \chi}{a_\perp \sin \chi} \ $. Inserting these values in (\ref{intercalottes0}), then inserting the integral in (\ref{EL_int}) and writing $|\chi|$ instead of $\chi$, we get the correlation coefficient:
\ba E_L(\vec a, \vec b) &=& 1 - \frac{2|\chi|}{\pi} + \frac{2}{\pi}a_z \arctan(\frac{a_\perp b_z - a_z b_\perp \cos \chi}{b_\perp \sin |\chi|}) \nonumber \\ && \qquad \qquad  + \frac{2}{\pi}b_z \arctan(\frac{a_z b_\perp - a_\perp b_z \cos \chi}{a_\perp \sin |\chi|}) \ . \nonumber \\ \ea

So far we have calculated this coefficient for settings in the north hemisphere of the Bloch sphere. If the settings are in the south hemisphere, one can use the fact that $E_L(\vec a,\vec b) = -E_L(\vec a,-\vec b) = -E_L(-\vec a,\vec b) = E_L(-\vec a,-\vec b)$. One can check that the above expression is actually still valid for all cases.

Note finally that in the above calculation, we didn't pay attention to particular cases, when the denominators in the fractions would be zero. For $\chi = 0$ or $\pi$, $E_L$ (as in eq (\ref{EL_chi})) can be obtained by taking the corresponding limit in the previous expression, or can be obtained directly in a much simpler way.

\subsection*{Appendix B: Allowed pairs ($\vec\lambda_a,\vec\lambda_b$) in our last proposal}
\label{App_constr_lambdaA_lambdaB}

Here we will argue that in our last proposal with two different local variables $\vec\lambda_a$ and $\vec\lambda_b$ for Alice and Bob, not all pairs ($\vec\lambda_a,\vec\lambda_b$) should be allowed.

The argument is based on the following observation: suppose that Alice measures along direction $\vec a = (a_\perp \cos \varphi_a, a_\perp \sin \varphi_a, a_z)$, and finds the outcome $\alpha=+1$; this projects Bob's state onto $\bra{+\vec a}\otimes\mathbbm{1} \ \ket{\psi(\theta)} = \sqrt{\frac{1+c a_z}{2}}\ket{\vec b_{\vec a}}$, with $\vec b_{\vec a} = (\frac{s a_\perp}{1+c a_z}\cos \varphi_a,-\frac{s a_\perp}{1+c a_z}\sin \varphi_a,\frac{c+a_z}{1+c a_z})$. If Bob then measures the setting $\vec b_{\vec a}$, he will necessarily get the result $\beta = +1$, and therefore
\ba P_Q(+-|\vec a,\vec b_{\vec a}) = 0 . \nonumber \ea
This in turn implies, for the EPR2 decomposition $P_Q = p_L P_L + (1-p_L)P_{NL}$ (with $p_L \neq 0$), that \ba P_L(+-|\vec a,\vec b_{\vec a}) = 0 . \nonumber \ea

This constraint must be satisfied by any setting $\vec a$ (which defines the setting $\vec b_{\vec a}$). To ensure this, we shall not allow pairs $(\vec\lambda_a,\vec\lambda_b)$ that may give the results $(\alpha=+1,\beta=-1)$, for some choice of settings of the form $(\vec a,\vec b_{\vec a})$.

\bigskip

To make this more explicit, let us fix the first local variable $\vec\lambda_a$. For simplicity, we assume that $\vec\lambda_a$ is in the $xz$ plane. If this is not the case, the analysis below would be slightly more tedious, but the final result would be the same.

\begin{figure}
\begin{center}
\includegraphics[width=0.47\columnwidth]{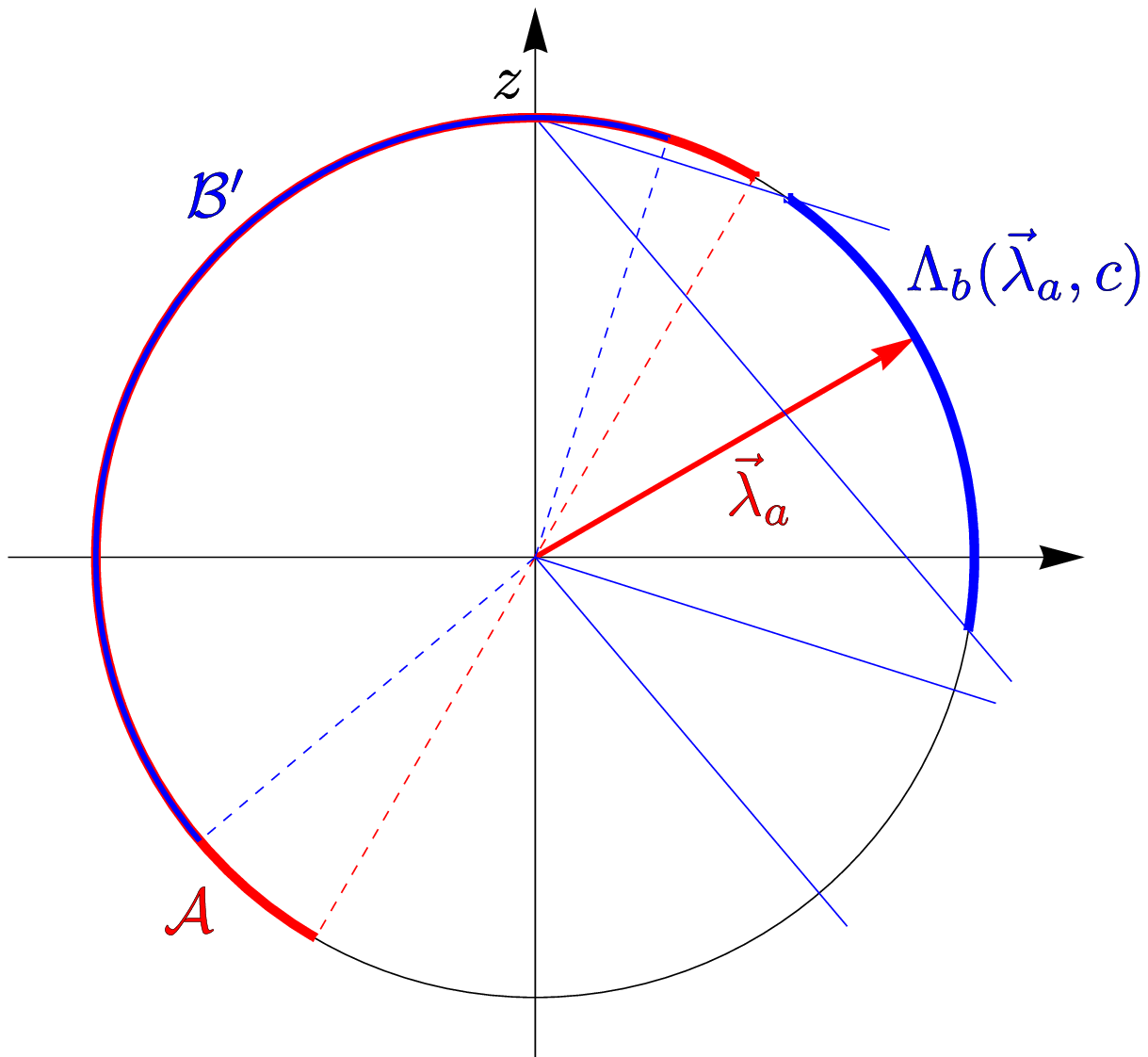}\hspace{0.2cm}
\includegraphics[width=0.47\columnwidth]{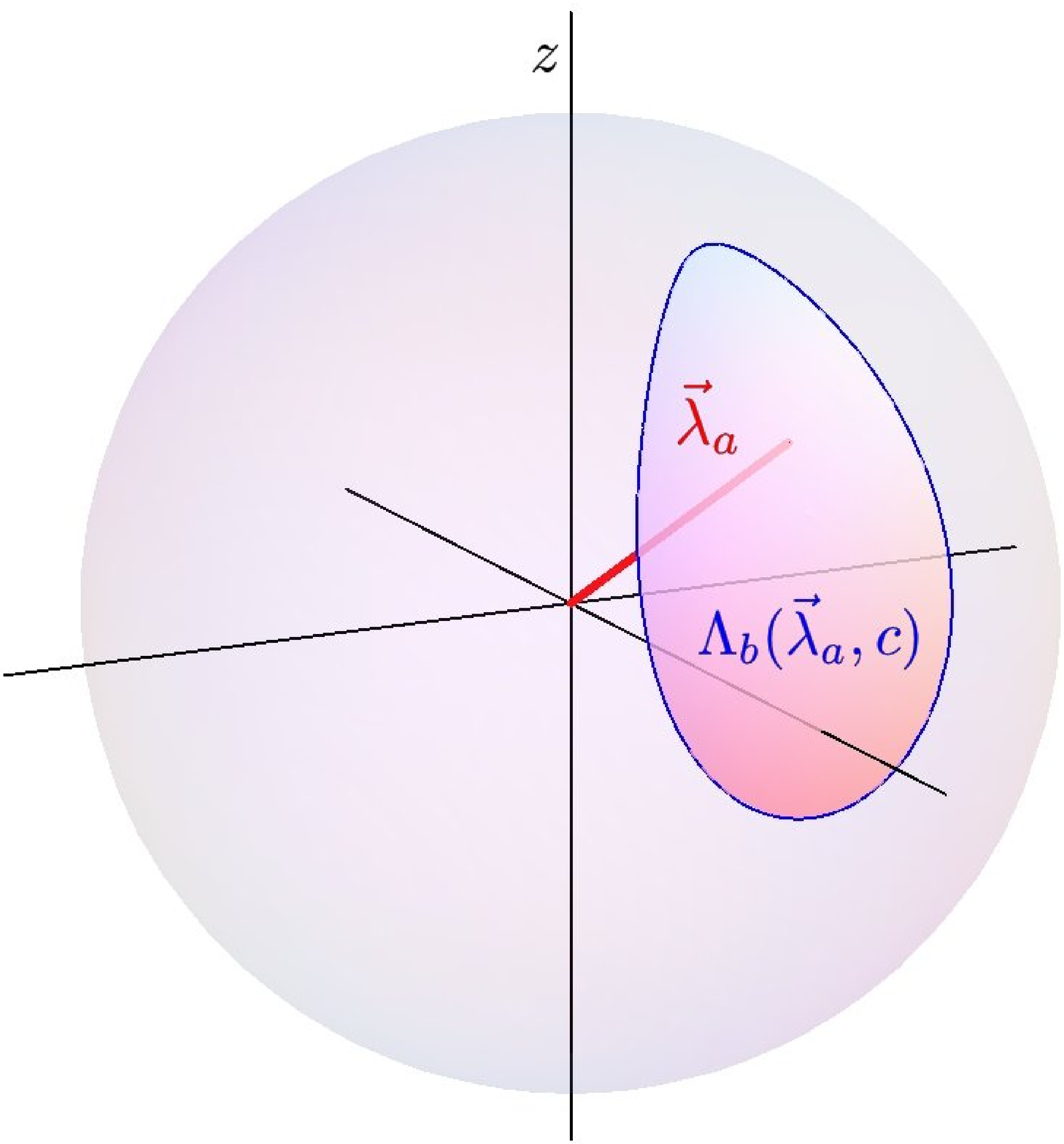}
\caption{Construction of the set $\Lambda_b(\vec\lambda_a,c)$ of allowed local variables $\vec\lambda_b$, given the first variable $\vec\lambda_a$. Left: 2D cut in the vertical plane that contains $\vec \lambda_a$. Right: 3D representation of $\Lambda_b(\vec\lambda_a,c)$. For both figures, $\vartheta_a = \frac{\pi}{3}$, $c=0.5$.} \label{fig_epr2_lambda_bs}
\end{center}
\end{figure}

The settings $\vec a$ that give the result $\alpha(\vec a,\vec \lambda_a)=+1$ span the half-sphere $\cal{A}$ above the bisector plane between $\vec z$ and $\vec\lambda_a$; see Figure~\ref{fig_epr2_lambda_bs} (left) for a 2D representation. $\cal{A}$ can be defined as \ba {\cal A} = \{ \vec a | \vec u \cdot \vec a \geq 0 \}, \quad \mathrm{where} \ \ \vec u = (-c_a, 0, s_a). \nonumber \ea
(Let us recall the notations: ($\vartheta_{a(b)}$,$\varphi_{a(b)}$) are the spherical coordinates of $\vec\lambda_{a(b)}$, and we write $c_a = \cos\frac{\vartheta_a}{2}, s_a = \sin\frac{\vartheta_a}{2}, c_b = \cos\frac{\vartheta_b}{2}, s_b = \sin\frac{\vartheta_b}{2}$ and $c_\varphi=\cos(\varphi_b-\varphi_a)$.)

The settings $\vec b_{\vec a}$, corresponding to these settings $\vec a \in \cal{A}$, then also span a spherical cap, $\cal{B}$, included in $\cal{A}$\footnote{Note: in particular, if $\vec\lambda_a$ is not assumed to be in the $xz$ plane, one would have $\cal{B}' \subset \cal{A}$ instead, where $\cal{B}'$ is the reflection of $\cal{B}$ with respect to the $xz$ plane.}. Using the fact that $a_z = \frac{(b_{\vec a})_z-c}{1-c (b_{\vec a})_z}$ and $a_\perp = \frac{s(b_{\vec a})_\perp}{1-c (b_{\vec a})_z}$, $\cal{B}$ can in turn be defined as
\ba & {\cal B} = \{ \vec b_{\vec a} | \vec u \cdot \vec a \geq 0 \} = \{ \vec b | -\vec v \cdot \vec b \geq c s_a \}, \nonumber \\ & \mathrm{where} \ \ \vec v = (s c_a, 0, -s_a). \nonumber \ea

According to the above observation, the allowed local variables $\vec\lambda_b$ must be such that for all those settings $\vec b_{\vec a}$ in $\cal{B}$, $\beta(\vec b_{\vec a},\vec \lambda_b) \neq -1$, i.e., $(\vec \lambda_b-\vec z) \cdot \vec b_{\vec a}' \leq 0$. This implies that $\vec \lambda_b-\vec z$ should be in a cone centered around $\vec v$, and with a half-angle $\xi = \arcsin \frac{c s_a}{||\vec v||}$. 
This writes
\ba \vec v \cdot \frac{\vec \lambda_b-\vec z}{||\vec \lambda_b-\vec z||} \geq ||\vec v|| \cos \xi = s \, . \nonumber \ea
For the fixed $\vec\lambda_a$ considered here, the set of allowed variables $\vec\lambda_b$ is then the intersection of this cone, translated by $\vec z$, and the Bloch sphere; see Figure~\ref{fig_epr2_lambda_bs}. Writing $\vec\lambda_b = (2 c_b s_b c_\varphi, 2 c_b s_b s_\varphi , 1-2s_b^2)$, the previous condition implies, that:
\be \mathrm{the \ pair \ } (\vec \lambda_a, \vec\lambda_b) \ \mathrm{should \ be \ allowed \ only \ if \ \ } \frac{s_a s_b}{1-c_a c_b c_\varphi} \geq s \ . \nonumber \ee
This justifies the constraint (\ref{constr_x}).

\newpage

\end{document}